\documentstyle[aps,prb,twocolumn,epsfig]{revtex}
\begin{document}
\draft
\title{Spin-dependent resonant tunneling
through semimetallic ErAs quantum wells
in a magnetic field}
\author{A. G. Petukhov}
\address{Physics Department, South Dakota School of Mines and Technology,
Rapid City, SD 57701-3995}
\author{W. R. L. Lambrecht and B. Segall}
\address{Department of Physics,
Case Western Reserve University, Cleveland, OH 44106-7079}
\date{\today}
\maketitle

\begin{abstract}
Resonant tunneling through semimetallic ErAs quantum wells embedded
in GaAs structures with AlAs barriers was recently found
to exhibit an intriguing behavior in magnetic fields:
a peak splitting occured only in fields perpendicular to the film while
a second resonant channel opened for in-plane fields.
The behavior is explained in terms of the valence bands states in
ErAs in the vicinity of the $\Gamma$ point,
their exchange splitting induced by the localized Er 4f magnetic moments,
and by a selection rule involving the total
angular momentum component along the normal to the interface.
\end{abstract}
\narrowtext

The utilization of thin metallic layers
as active components in semiconductor devices  such as metal base
and resonant tunneling transistors opens new avenues in micro-electronics.
Resonant tunneling through thin metallic layers
embedded in a semiconductor has been reported by several
authors.\cite{Tabatabaie}
A promising materials system for the study of this phenomenon
is semimetallic ErAs embedded in GaAs
since the two are closely lattice matched.\cite{Palmstrom}
Because of the open 4f shell of Erbium, these heterostructures also have
interesting magnetic
behavior.\cite{Allenprl,Allenprb,Bogaerts,Petukhov,Petukhovb}
Here we provide a theoretical explanation
for the unusual behavior of resonant tunneling
through ErAs layers in  magnetic fields of different orientation
which was recently reported by Brehmer {\sl et al.}\cite{Brehmer}
and Zhang et al..\cite{Zhang} These experiments were carried out on
resonant tunneling diode (RTD) structures
consisting of ErAs quantum wells and AlAs barriers
sandwiched between the $n+$ doped GaAs substrate and GaAs capping layers.

The measurements of differential conductance
($dI/dV$ vs. $V$) \cite{Brehmer}, reproduced
in Fig. \ref{figexpt}, indicate the presence of
two different resonant tunneling channels called A and B.
The peak A
splits in a magnetic field perpendicular to the layers, but remains
nearly unaffected (slightly shifted)
by a magnetic field parallel to the layers. The B channel,
on the other hand, shows no observable splitting for either
direction of the field, is only weakly resolved as a shoulder
in zero or perpendicular field, but is strongly enhanced by a
field parallel to the film.
The most extensive results and the ones we focus on here were obtained
for the [113] orientation of the film. However, similar results
were obtained for other orientations of the films.
That indicates that the phenomena are not determined by specific
intrinsic crystallographic directions but rather by the relative orientation
of field and interface.

In the present letter, we will show that this behavior is explained
by tunneling of electrons into the ErAs light  and heavy hole states
in the vicinity of the $\Gamma$-point for the A and B resonances respectively.
In the absence of a magnetic field or in  a perpendicular field, tunneling is
predominantly only into the light hole bands because of an approximate
conservation
of the total angular (orbital and spin) momentum component
along the normal of the film. The exchange splitting of these states
induced by coupling to the open 4f shell leads to an anomalously large
Zeeman splitting. On the other hand, a parallel field allows
for tunneling into the heavy hole band  leading to an enhancement of
channel B.  The splitting in this case will be shown to be reduced.

To understand the tunneling
behavior, a detailed understanding of the ErAs
band structure and magnetic properties is required.
This is provided by our recent electronic structure calculations
\cite{Petukhov,Petukhovb} of ErAs and related Er$_x$Sc$_{1-x}$As
alloys, the results of which were in good agreement  with magnetotransport
measurements.\cite{Bogaerts}
The first question to be addressed
is whether the tunneling takes place into the
conduction (electron) or valence (hole) states. This is expected
to depend on the film direction. For the [113] direction under
consideration, the band structure (see Fig. \ref{fig113})
shows clearly that unoccupied electron states are more than 1.5 eV
above
the Fermi level.
Hence they cannot be involved in the tunneling.
In fact, although the Schottky barrier height at the
ErAs/AlAs interface is presently unknown, the Fermi level in the
ErAs must line up with the
donor states near the conduction band edge deep in the $n+$ doped GaAs
region. Implicit in the above argument is that $k_\parallel$ is conserved
in the resonant tunneling and that only electrons with $k_\parallel\approx0$
are involved in the transport from GaAs
because the conduction band minimum in GaAs is at $\Gamma$.

Using the simple infinite barrier model, a first estimate can be obtained
for the quantum confined states in the ErAs quantum well.
For a film thickness of $N$ monolayers (ML), there are  $n=1 \dots N$
quantized values
of wave vector $k_{\perp,n}=n \pi/Na_\perp$, where $a_{\perp}$ is the
distance between two monolayers.
The corresponding values of $k_\perp$ are indicated as vertical lines
in Fig. \ref{fig113} for a layer thickness of 12 ML.
This shows that the only the $n=1$ quantized hole states lie close to
the Fermi level and are likely to be involved in the tunneling.

To relate these quantized energies $E_n$
to the voltages $V$ at which resonance
occurs, we need to know the voltage profile in the RTD.
This is somewhat uncertain because the top and bottom interfaces
of ErAs are of different quality due to difficulties in the epitaxial
overgrowth on ErAs. For a symmetric structure, there would be equal voltage
drops across
the two insulating GaAs space charge layers
and AlAs barriers leading to a
resonance when $V/2\approx E_n$.
Direct measurements of the voltage drop on three terminal
devices, however have shown that most of the voltage drop occurs
on the substrate to ErAs interface.\cite{Brehmer}  This implies that
$E_n\approx V$.

Unlike our earlier calculations \cite{Petukhov},
the band structure shown in Fig. \ref{fig113} includes the effect of
spin-orbit coupling which, as will be shown below, is essential for
understanding the magnetic field dependence.  Otherwise, the details
of the calculation method are similar to those of Ref. \cite{Petukhov}.
The calculations are performed within the atomic sphere approximation
of the linear muffin-tin orbital method\cite{Andersen} and are based
on local spin density functional theory. The spin-orbit coupling
terms are added as a perturbation to the spin-dependent Hamiltonian,
which is then diagonalized numerically without further
approximations.\cite{Andersen}
The spin-polarization arises from the alignment of unpaired spins
in the 4f level which is treated as an open shell core-like level
without dispersion. In the absence of a magnetic field, the 4f
localized spin magnetic moments
are randomly oriented and the bands are described by a calculation
without spin-polarization.

The manifold of valence
states around $\Gamma$ can be described by the Kohn-Luttinger (KL)
Hamiltonian,
\cite{Luttinger} which accounts for the behavior up
to quadratic terms in $\bf k$. In the following,
we use this Hamiltonian in the spherical approximation
(i.e. $\gamma_2=\gamma_3$ in terms of the conventional
Luttinger parameters \cite{Luttinger})
in order to gain a qualitative understanding of the
magnetic field effects. We stress, however, that this
description of the nature of the quantum states involved in the tunneling
and the associated conservation laws remain approximately valid
even in the complete calculation. The KL-Hamiltonian we start from is:
\begin{eqnarray}
H =&-&(\gamma_1 + 4\gamma_2){\bf k}^2 + 6\gamma_2({\bf L}\cdot{\bf k})^2
\nonumber \\
 &+& (\lambda_0 +
\lambda_1{\bf k}^2 + \lambda_2({\bf L}\cdot{\bf k})^2)(
{\bf S}\cdot\hat{\bf B}) \nonumber \\  &+&
\frac{2}{3}\Delta_{SO}({\bf L}\cdot{\bf S} - \frac{1}{2}),
\label{ham}
\end{eqnarray}
where $\bf L$ is the $l=1$ orbital momentum operator describing p-like
hole states at the top of the valence band of ErAs, $\bf S$ is the spin
operator and $\hat{\bf B}$ a unit vector along the magnetic field.
The first two terms are the usual Luttinger terms.\cite{Luttinger}
The last term is the spin-orbit coupling term
with the $1/2$ inside the parenthesis being
a reference-level shift to the valence band maximum.
The parameters $\lambda_i$ describe the splitting of the valence
states in the presence of a magnetic field.
Here the splitting
is primarily due to the exchange interaction of the holes
with the ferromagnetically aligned localized moments of 4f electrons.
The alignment of the latter is well-described by standard Brillouin theory
of paramagnets.\cite{Allenprb}
In our numerical calculations we assume saturated
spin-polarization. This is valid because
the measurements were taken at $T=4$K and the applied
magnetic field was 8 T.\cite{Allenprb}
Direct coupling of the angular momentum
to the magnetic field (the Zeeman effect)
is neglected as it is much smaller
than the coupling to the exchange field.
Also neglected are the terms that lead to the formation of
Landau levels. The above Hamiltonian is thus equivalent to
the one used in our first-principles calculations and its parameters
can be extracted from the latter.

For reasons already explained above, we can now set
$k_\parallel=0$ and choose the $z$ axis along the normal to the interface.
The quantization of $k_\perp$ in the infinite barrier model
then fixes the $\bf k$ completely.
In the absence of the spin-orbit interaction the above
Hamiltonian (\ref{ham}) describes two
sets of spin-polarized bands that are independent of the direction of the
magnetic field.
The spin-orbit interaction now couples the spin-direction (determined
by the magnetic field) to the orbital angular momentum direction (fixed by
$k_\perp$ to be normal to the interface). This is what leads to the non-trivial
magnetic field orientation dependence.
More precisely, without the magnetic field term but
including spin-orbit interaction the Hamiltonian commutes
with $J_z$ where $J_z=L_z+S_z$ is the projection of the
total angular momentum onto the direction normal to the interface.
If the magnetic field is along the same direction
(${\bf B} \parallel \hat{\bf z}$),
the Hamiltonian (\ref{ham}) maintains this cylindrical symmetry
and splits up into two $2\times2$ submatrices $H_{n,\pm{1/2}}^{\perp} - E_n^0$,
describing the coupling between the $|3/2,\pm1/2\rangle$
and $|1/2,\pm1/2\rangle$ states,

\begin{equation}
\left(
\begin{array}{cc}
 \frac{1}{6}(2A_n \pm B_n \mp C_n) &
 \frac{\sqrt{2}}{6}(2A_n \mp 2B_n \mp C_n)\\
 \frac{\sqrt{2}}{6}(2A_n \mp 2B_n \mp C_n) &
 \frac{1}{6}(4A_n \pm B_n \pm 2C_n)
 - \Delta_{SO}
\end{array} \; \; \right) \;,
 \label{matperp}
\end{equation}
where $E_n^0 = -(\gamma_1+4\gamma_2)k_{\perp,n}^2$,
$A_n=6\gamma_2k_{\perp,n}^2$,
$B_n=\lambda_0+\lambda_1k_{\perp,n}^2$, and $C_n=\lambda_2k_{\perp,n}^2$.
We will primarily be concerned with $n=1$ states.
Similarly the eigenvalues of the $|3/2,\pm3/2\rangle$ states are
\begin{equation}
E_{n,\pm3/2}=E_n^0 + A_n \pm D_n/2,
\label{eigen3/2}
\end{equation}
where $D_n=B_n+C_n$.

In the case of the magnetic field parallel to the interface,
$J_z$ no longer commutes with the Hamiltonian
and there is an additional mixing of the eigenstates of the original
Hamiltonian (without field).  The
energy spectrum of the quantum well can, however, be found
as the eigenvalues
of the two $3\times3$ submatrices
$H_{n,\pm{}}^{\parallel} - E_n^0$,
written in the basis
$\frac{1}{\sqrt{2}}(|3/2,3/2\rangle \pm |3/2,-3/2\rangle)$,
$\frac{1}{\sqrt{2}}(|3/2,1/2\rangle
\pm |3/2,-1/2\rangle)$, and $\frac{1}{\sqrt{2}}(|1/2,1/2\rangle
\mp |1/2,-1/2\rangle)$

\begin{equation}
\left(
\begin{array}{ccc}
A_n & \pm\frac{1}{\sqrt{12}}D_n & \pm\frac{1}{\sqrt{6}}D_n \\
\pm\frac{1}{\sqrt{12}}D_n & \frac{1}{3}(A_n \pm B_n) &
\frac{\sqrt{2}}{6}(2A_n \mp{B_n})\\
\pm\frac{1}{\sqrt{6}}(D_n) & \frac{\sqrt{2}}{6}(2A_n \mp{B_n}) &
\frac{1}{6}(4A_n \pm B_n) - \Delta_{SO}\\
\end{array} \; \; \right).
\label{matpar}
\end{equation}

The analysis of the structure of
the matrices (\ref{matperp}-\ref{matpar}) allows one to explain
semiquantitatively all the features of the resonant tunneling
experiments in a magnetic field.
First, we recall that the electrons tunneling form the $n^+$ GaAs
conduction band minimum are $s$-like and have $M_J=\pm1/2$.
With the magnetic field absent or
perpendicular to the layers,  $M_J$ is a good quantum number that
must be (approximately) conserved. Hence
tunneling transitions between states with $|M_J|=1/2$ and
$|M_J|=3/2$ are forbidden.
This explains why in that case only the A-channel (corresponding to
tunneling into the light hole states with $M_J=\pm1/2$)
is pronounced. The small shoulder of the B-channel is due to the fact
that in the real band-structure the axial symmetry is slightly broken.
For parallel fields, the $M_J$ states are mixed and the B-channel becomes
available for tunneling.
The observed splitting in the perpendicular magnetic field  is also
easily understood since the energies depend on the
sign of $M_J$.  The magnetic field dependence of this splitting was shown
by  Zhang et al.\cite{Zhang} to follow the Brillouin theory of
paramagnetic alignment of localized spins as expected in the present theory.

Fig. \ref{figspher}(a,b)
shows the eigenvalues of the above matrices in perpendicular and in-plane
field respectively, with the parameters  $E_n^0=-84.3 k_\perp^2$,
$A_n=59.7 k_\perp^2$,
$B_n=-9.75 k_\perp^2$, $C_n=14 k_\perp^2$, with $k_\perp$
in units $\pi/a_\perp$ and $\Delta_{SO}=0.45$ and all energies in eV.
These parameters were extracted
from our first-principles calculations and appropriately adjusted for the
spherical approximation.
The small spin-splitting at $\Gamma$, $\lambda_0\approx1$ meV
was neglected. As discussed in our previous work \cite{Petukhov},
the Fermi-level  position was slightly shifted
from its LDA value as it was found that the LDA overestimates the
Fermi surface volume \cite{Petukhov} because of the neglect of
electron interaction self-energy effects.
This figure  clearly shows that the light-hole spin-splitting
is about 5 times larger for the perpendicular than for the in-plane field.
Also indicated in this figure are the experimental resonance positions
for varying thicknesses (i.e. different $k_{\perp,n}$) assuming $V=E_n$.
Overall, the agreement is quite good for the absolute
energies of the resonances, the dispersions, and
the light-hole heavy hole splitting. The spin-splitting appears to be
slightly underestimated in the calculation. Experimentally the  splitting
was seen to be about 80 meV, while the model gives 50 meV.

The advantage of the present analytical model is that we can easily
investigate the effect of varying parameters. Because
spin-orbit interaction is fairly large,
the coupling to the spin-orbit split-off
bands can be treated in perturbation theory.
Up to first order terms in $1/\Delta_{SO}$,
the light hole spin splitting in perpendicular field  is:
$\Delta_{xc}^\perp(lh)\approx (B_n-C_n)/3-4A_n(2B_n+C_n)/9\Delta_{SO}$,
which is of order 50 meV,
while for the in-plane field, it is:
$\Delta_{xc}^\parallel(lh)\approx 2B_n/3-4A_nB_n/9\Delta_{SO}$,
which is of order 10 meV.
Similarly, we find that the heavy hole splitting for the in-plane
field up to terms in $1/\Delta_{SO}$ is zero, which explains why
no spin-splitting is observed for the B-channel. For the perpendicular
field, the heavy hole splitting is $B_n+C_n\approx40$ meV but
is not observed because in that case this channel is reduced
to a weak shoulder on a rising background.
Again up to the same order in $1/\Delta_{SO}$,
the component of the  allowed $J=1/2$ channel in
the light hole eigenstate for in-plane field is
given by $\{1-[(D_n/\sqrt{12})(1+2A_n/\Delta_{SO})/(2A_n/3)]^2\}^{1/2}$
which is about 94\%.  This is consistent with the fact that
the A-channel is not noticeably reduced in intensity when
the in-plane field is switched on.

A quantitative calculation
of resonance intensity for the B channel due to the weak
breaking of axial symmetry and its gradual switching on in  the magnetic
in-plane field will require a more detailed study of the quantum tunneling
process. Finally, we note that the resonances in this system are relatively
broad compared to typical resonant tunneling in semiconductor quantum wells.
Among other broadening mechanisms, this may result from the loosening
of $k_\parallel$ conservation due to the
roughness of the top layer interface. We thus anticipate that
with further improvements in the material some of the finer details
of our model may be checked experimentally in future work.

We have also carried out fully first-principles calculations of the
bands in the [113] direction in parallel and perpendicular magnetic field
including the spin-orbit coupling. These calculations qualitatively confirm
the above results, but lead to a somewhat smaller
light hole spin splitting (30 meV) in perpendicular field
than the spherical model (which was already too small)
and slightly lower masses
and hence slightly lower level positions for a given layer thickness.
A possible origin for this discrepancy is that our present
calculations are based entirely on bulk band structures.
In the quantum wells,
mixing of Er 5d derived bands, which have larger
spin-splitting, into the relevant hole states will occur
and may lead to a substantial enhancement of the
exchange splitting effects.

In conclusion, the spherical Kohn-Luttinger model  of the
valence band maximum supplemented with a strongly enhanced spin Zeeman effect
(due to coupling of the valence electrons with the localized
magnetic moments of the 4f electrons which are aligned
in the field)  provides a good semiquantitative
description  of the ErAs valence band maximum and its associated
eigenstates. The tunneling behavior can be understood in terms of
approximate conservation of the total angular momentum component along the
normal to the interface. For perpendicular or zero field this forbids the
transitions into the heavy hole bands while the latter become allowed
by mixing of states for an in-plane field.
The origin of this approximate selection rule is that the
$k_\parallel\approx0$
of the electrons originating in the GaAs conduction band is
conserved. This  in turn fixes the quantization axis for the orbital
angular momentum along the normal to the interface. The spin direction
on the other hand is fixed by
the magnetic field. The spin-orbit coupling is thus
essentially responsible for the qualitatively different behavior
of the tunneling in the parallel and perpendicular fields.

We wish to thank S. J. Allen and D. E. Brehmer for stimulating discussions
and for providing their experimental data before publishing.
This work was supported by  the Air Force Office of Scientific Research
Grant No. F49620-95-1-0043 and by the National Science Foundation under Grant
No. OSR-9108773 and the South Dakota Future Fund.

\begin{figure}
\vskip 4 cm
\caption{Differential conductance of 12.8 ML ErAs
resonant tunneling diode in perpendicular and in-plane magnetic
fields from Ref. \protect{\cite{Brehmer}}. The A- and B-resonant
channels are indicated.\label{figexpt}}
\end{figure}

\begin{figure}
\begin{center}
\mbox{\epsfig{file=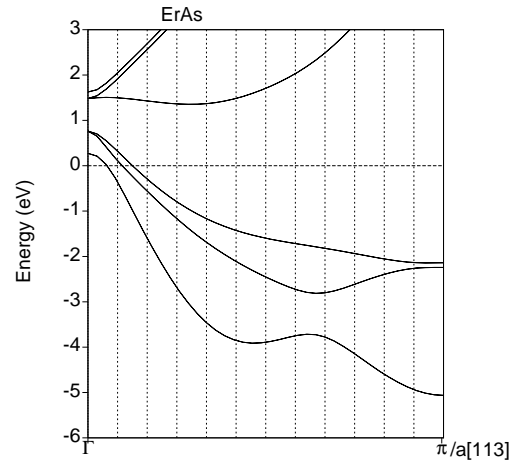,width=6cm}}
\end{center}
\caption{Energy band structure of ErAs along the [113]
direction in zero field. The vertical lines indicate
quantized $k_{\perp,n}$ values in an infinite barrier
quantum well of 12 ML.\label{fig113}}
\end{figure}

\begin{figure}
\begin{center}
\mbox{\epsfig{file=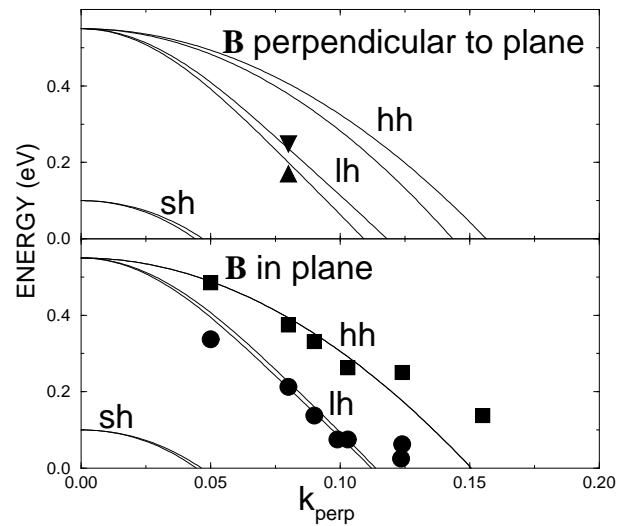,width=8cm}}
\end{center}
\caption{Band structure of ErAs near the valence band maximum
in the [113] direction in a saturating magnetic field which
in (a) is perpendicular and in (b) is parallel to the (113) plane
within the spherical Kohn-Luttinger model.
Experimental resonance positions given by $V=E_n$ (see text)
for $n=1$ and
different film thicknesses (or $k_{\perp,n}$ values) are indicated
by circles for the A-channel and squares for the B-channel.
The experimental spin-split A-channel resonances are indicated by
triangles for one layer thickness in (a). The heavy hole (hh),
light hole (lh) and split-off hole (sh) bands are labeled.
\label{figspher}}
\end{figure}

\end{document}